\documentclass{iopart}
\usepackage[latin1]{inputenc} 
\usepackage{color}
\usepackage{amssymb}
\expandafter\let\csname equation*\endcsname\relax
\expandafter\let\csname endequation*\endcsname\relax
\usepackage{amsmath}
\usepackage{color}
\usepackage{dsfont}
\usepackage{graphicx}
\usepackage{picinpar}
\usepackage{wrapfig}
\usepackage{movie15}
\usepackage{dsfont}
\usepackage{float}
\usepackage{paralist}
\usepackage{hyperref}
\usepackage[format=hang,margin=10pt]{subfig}

\newcommand{\abs}[1]{\ensuremath{\left| #1 \right|}}
\newcommand{\ve}{\ensuremath{\varepsilon}}
\newcommand{\fsust}{\ensuremath{f_\text{sust}}}

\usepackage{color} 
\newcounter{commentzaehler}

\begin{document}

\title{Effect of small-world topology on wave propagation on networks of excitable elements.}
\author{T. Isele and E. Sch\"oll}
\address{Institut f\"ur Theoretische Physik, Technische Universit\"at Berlin,
10623 Berlin, Germany}
\eads{schoell@physik.tu-berlin.de}

\begin{abstract}
We study excitation waves on a Newman-Watts small-world network model of coupled excitable elements.
Depending on the global coupling strength, we find differing resilience to the added long-range links and different mechanisms of propagation failure.
For high coupling strengths, we show agreement between the network and a reaction-diffusion model with additional mean-field term.
Employing this approximation, we are able to estimate the critical density of long-range links for propagation failure.
\end{abstract}

\maketitle


\pagestyle{plain}
\setcounter{page}{1}
\pagenumbering{arabic}
\section{Introduction}
Excitable media are well studied model systems in a variety of applications ranging from chemical \cite{MIK94} to electronic systems \cite{SCH01} and lasers \cite{SOR13} and from heart-muscle tissue \cite{KEE98} to neural systems \cite{IZH07,DAH09a}.
An excitable system rests in a stable steady state, but after a sufficiently strong perturbation performs a long excursion in phase space, i.e., emits a spike, before returning to the stable steady state again.
Excitable media arise when excitable elements are coupled spatially.

The spatial coupling facilitates an abundance of dynamical behavior amongst which are Turing patterns \cite{TUR52}, traveling waves \cite{MIK94}, spots \cite{KRI94} and spiral waves \cite{DAV92a}, to name just a few.
Especially wave-like spatio-temporal behavior is interesting from a neuroscience point of view.
It is observed in living neural tissue and considered to play a role in neural information processing in different tasks \cite{RUB06,SAT12}.
Traveling waves are a generic phenomenon in cortical dynamics \cite{MUL12}, and have successfully been used to describe features of cortical spreading depolarization \cite{DAH08,DAH12b,KNE14}.

In recent years dynamical systems coupled in complex network architectures have attracted a lot of attention \cite{BOC06a,NEW03,ALB02a,DAH12,WIL13}.
Those systems can also occur in a wide variety of applications ranging from power grids \cite{ROH12} to  biological networks \cite{MON06a}.
In neuroscience,  scenarios with excitable elements coupled in a chain-like one-dimensional topology have been suggested as a mechanism for the occurrence of traveling waves of activity in the visual cortex \cite{SAT12}.
In related works, a one-dimensional network model of pyramidal cells and interneurons produces saltatory propagation, and excitatory connections play a crucial role \cite{MUL12}.

The topology of a network is a key factor influencing the dynamic behavior of the system.
Network topologies can range from well-ordered lattice systems, which resemble spatially extended systems, to random topologies where the notion of space loses its meaning.
A very interesting intermediate form are the so-called small-world topologies \cite{MIL67,WAT98}, in which strong local connectivity is combined with a few long-range connections enabling short mean path lengths.

Small-world topologies are of interest for the description of anatomic and functional brain networks \cite{BAS06a}.  
They are considered a powerful and versatile approach to the structure of those systems, and it is argued that within the cerebral architecture, they are preferred as a trade-off between network efficiency and wiring cost \cite{BUL12}.

On the dynamical side, wave and front propagation have been studied in network topologies such as trees, where fronts can be pinned and waves cease to propagate \cite{KOU12,KOU14}.
Small-world topologies have been shown to make systems support sustained activity \cite{ROX04,SIN07b}.

Regarding the influence of topology on the behavior of traveling waves,
it has been shown that short-range connections mediate traveling waves in phase oscillator models which are used as simple models for cortical waves \cite{ERM01a}.
Also, additional long-range connections in a chain of locally coupled oscillators can be used to generate traveling waves of different wavelengths \cite{ERM94}.

A change of topology may occur in pathological states as, e.g., multiple sclerosis, which affects the most expensive (long-range) links in the brain network \cite{BUL12}.
Moreover, propagation failure of waves is an important aspect in different areas of physiology \cite{DEL87,KEE09}.

To the best of our knowledge, the combination of these relevant ingredients, i.e., generic excitable elements, small-world topology and propagation failure, has not been addressed in a satisfactory way, except for very few exemptions, e.g.~\cite{VAN04}.
We attempt to do so by using a generic model of excitability, the well-known FitzHugh-Nagumo model \cite{FIT61,NAG62}, combined with a Newman-Watts small-world architecture \cite{NEW03} as well as techniques from spatially continuous systems to study the behavior of excitation waves.

The structure of the paper is as follows:
In section~\ref{sec:model}, we introduce the dynamics and the network model.
We draw a connection between a ring network and a one-dimensional continuous reaction-diffusion system. 
We discuss traveling wave solutions in both systems, in particular their spectral and stability properties, and we elaborate on the difference between the dynamics of the network and the reaction-diffusion system.
In section~\ref{sec:wave_on_sw}, we consider a small-world topology. We highlight the different mechanisms leading to propagation failure in different dynamical regimes. 
In section~\ref{sec:analytic}, we modify the continuous reaction-diffusion model in order to incorporate the effect of the small-world topology. 
We give an estimate of the critical link density at which excitation waves cease to exist in the small-world network.
In section~\ref{sec:conclusion}, we summarize our findings.

\section{Model}\label{sec:model}
\subsection{Dynamics} \label{sec:dynamics}
As a generic model of excitable dynamics we use the FitzHugh-Nagumo model \cite{FIT61,NAG62} on an undirected, unweighted  network, where neighboring nodes are coupled by the difference in the activator concentrations.
Thus, the model reads
\begin{subequations}\label{eq:model}
\begin{align}
  \dot{u}_i  &= u_i-\frac{u_i^3}{3} -v_i + D \sum_{j=1}^N \mathcal{A}_{ij} (u_j-u_i)             \label{eq:model_u}\\
  \dot{v}_i  &= \ve (u_i + \beta),  \quad i=1,...,N,                                           \label{eq:model_v}
\end{align}
\end{subequations}
where $u_i$ is the activator, and $v_i$ the inhibitor at node $i$, and $\mathcal{A}_{ij}$ is the adjacency matrix of the network, $D>0$ is the coupling strength, 
$\ve\, (\ll 1)$ and $\beta$ are the time-scale separation and the excitation threshold of the local dynamics, respectively. 
The local dynamics of Eq.~\eqref{eq:model} (i.e., without the coupling term) possesses a steady state at $u^*=-\beta,\ v^*=-\beta+\beta^3 /3$.  
For a value of $\abs\beta>1$, this steady state is stable. 
It undergoes a supercritical Hopf bifurcation at $\abs\beta=1$.
For the network Eq.~\eqref{eq:model} a linear stability analysis shows that the eigenvalues of the linearization around the homogeneous steady state $u^*_i = u^*,\ v^*_i=v^*$ are given by $\mu^{\pm}_j = \frac 1 2 \left( 1-u^{*2} + D \lambda_j \pm \sqrt{\left(1-u^{*2}+D\lambda_j\right)^2-4\ve}\right)$, where $\lambda_j$ are the eigenvalues of the Laplacian matrix $\mathcal{L}_{ij}=\mathcal{A}_{ij}-\sum_{k=1}^N \mathcal{A}_{ik}\delta_{ij}$ of the network, $\lambda_j\le 0$, with at least one eigenvalue $\lambda_i=0$.
For $\lambda_i=0$, $\mu^{\pm}_i$ are the eigenvalues of the local dynamics. 
Comparing terms of $\mu^{\pm}_j$, we note that the real part of the square root term is always smaller than the absolute value of the term before the square root.
Thus, when the former is negative, $\mu^{\pm}_j$ is also negative.
This is always the case if $u^{*}>1$ and thus if the steady state of the local dynamics is stable.
On the other hand, if the term before the square root is positive, then $\mu^{+}_j>0$.
A sufficient condition for this is  $u^{*}<1$ and $\lambda_j=0$, which, as there always exists a $\lambda_j=0$, always happens if the steady state of the local dynamics is unstable. 
Thus, we conclude that the homogeneous steady state inherits the stability of the local equations and thus when $u^*,v^*$ is stable, so is the homogeneous steady state $u_i \equiv u^*,\ v_i\equiv v^*$.

\begin{figure}[tb!]
\subfloat{%
\includegraphics[width=0.5\textwidth]{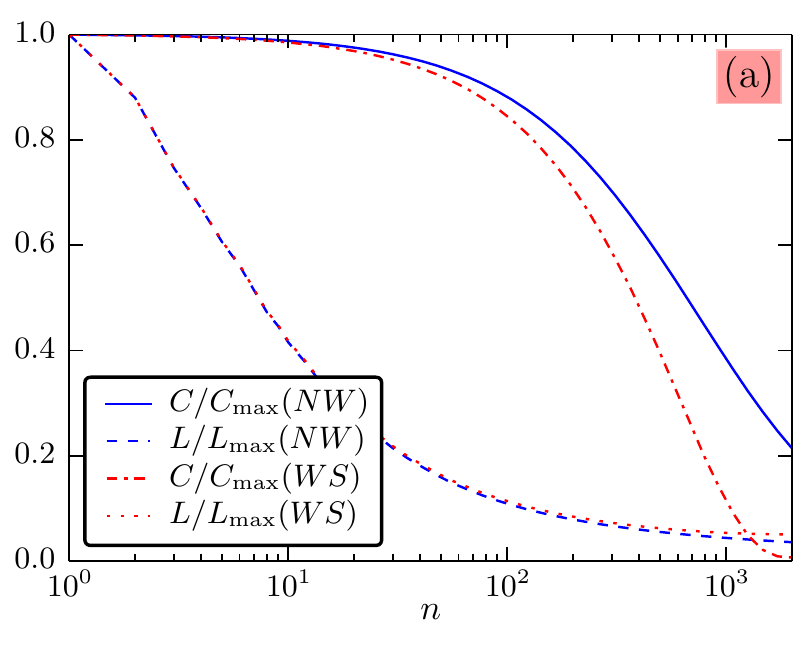}}%
\subfloat{%
\includegraphics[width=0.5\textwidth]{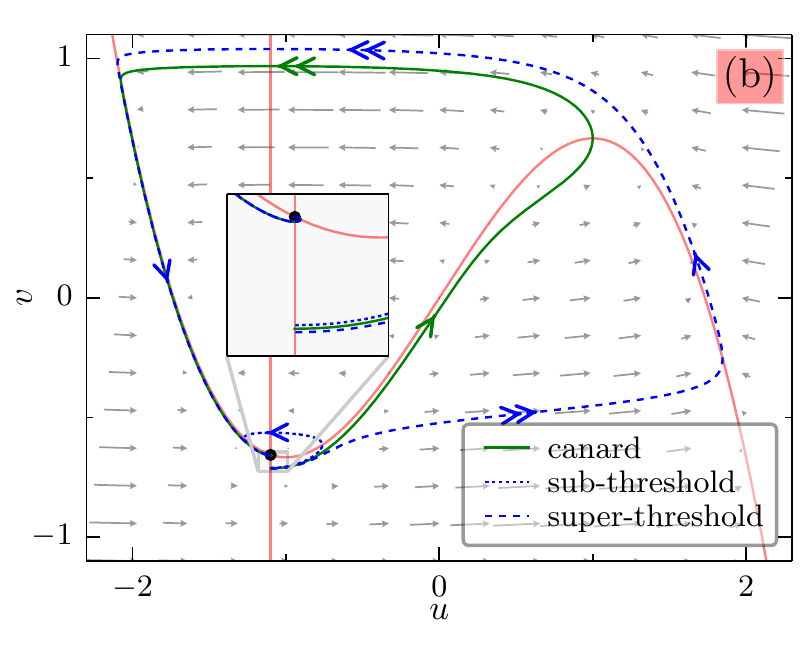}}%
\caption{(a) Normalized (global) clustering coefficient $C$ and average shortest path length $L$  vs. number of additional random links $n$ for the Watts-Strogatz (WS) (red dash-dotted, dotted) and Newman-Watts (NW) small-world model (blue solid, dashed). 
Parameters: $N=1000$ and $R=2$. 
(b) Phase portrait of ($u$,$v$) in the FitzHugh-Nagumo system: nullclines (red solid), canard trajectory (green solid), and two solutions with initial conditions slightly below (blue dotted) and slightly above (blue dashed) the threshold given by the canard trajectory.  The inset shows a blow-up near the steady state. 
Parameters: $\beta=1.1$, $\ve=0.04$
\label{fig:nw_model_and_phaseplane}
}
\end{figure}

For $\ve\ll 1$, the local dynamics of Eqs.~\eqref{eq:model} is a prototypical example for a slow-fast system showing type-II excitable behavior.
Systems of type-II excitability do not possess a constant threshold that separates stimulations leading to a spike from those that do not. 
They are rather characterized by a ``threshold trajectory'', around which the system is very sensitive to the size of the stimulus.
In the case of the FitzHugh-Nagumo system one commonly chooses as threshold trajectory the so-called `canard trajectory' that goes through the $v$-maximum of the $u$-nullcline. It is marked by a green solid line in Fig.~\ref{fig:nw_model_and_phaseplane}(b).
Also shown in this Figure are one trajectory with initial conditions slightly above this threshold trajectory and one slightly below it, leading to sub- (blue, dotted) and super-threshold (blue, dashed) behavior, respectively.
For a more detailed account on excitability in the FitzHugh-Nagumo system see the supplementary material and \cite{IZH07}.
In the remainder of this work, we will fix the parameters at $\ve=0.04$ and $\beta=1.1$, such that the model is in the excitable regime.

\subsection{Traveling wave solutions}\label{sec:traveling-waves}
As a preliminary study, we will discuss the behavior of traveling wave solutions on a regular ring network.
Consider a ring topology with $N$ nodes, where each node is coupled to its $R$ neighbors to the left and its $R$ neighbors to the right, so that every node has degree $2R$.
Equations \eqref{eq:model} on such a system read
\begin{subequations}\label{eq:ring_system}
\begin{align}
  \dot{u}_i  &= u_i-\frac{u_i^3}{3} -v_i + D \sum_{r=1}^R (u_{i-r} + u_{i+r} -2u_i)         \label{eq:ring_system_u} \\
  \dot{v}_i  &= \ve (u_i + \beta)\,,                                                       \label{eq:ring_system_v}
\end{align}
\end{subequations}
where $i=1,...,N$ and all indices are to be understood modulo $N$.

Apart from the stable homogeneous steady state discussed above, this system also supports traveling wave solutions.
At this point, we are interested in traveling wave solutions on the ring network in which exactly one region of excitation travels around the ring in either clockwise or counter-clockwise direction.
It turns out that the coupling strength $D$ in Eqs.~\eqref{eq:ring_system} has a significant influence on the existence, speed, and stability of these solutions.
At high coupling strengths $D$, the system Eqs.~\eqref{eq:ring_system} behaves much like a continuum reaction-diffusion system, whereas at low coupling strengths $D$, the discrete nature of Eqs.~\eqref{eq:ring_system} becomes important.
In the following, we will briefly discuss the behavior of these traveling wave solutions with changing coupling strength $D$.
We start by introducing the limiting continuum system.

\subsubsection{Traveling waves in the continuum limit.}\label{sec:trav-waves-cont-lim}
To examine the behavior of traveling wave solutions of Eqs.~\eqref{eq:ring_system} at large coupling strengths $D$, we define a continuum limit.
If we define a continuous spatial variable $x$ and a distance $h\equiv\frac 1{\sqrt{D}}$ between two adjacent nodes on the ring and assume that there is a function $u(t,x)$ such that $u_i(t)\equiv u(t,\frac{i}{\sqrt{D}})$, Eq.~\eqref{eq:ring_system_u} can be expressed as
\begin{align}
 \partial_t u &= u-\frac{u^3}{3}-v + D\sum_{j=1}^R\frac{u(t,x-jh)+u(t,x+jh)-2u(t,x)}{h^2}\,.\label{eq:ring_intermediate_step}
\end{align}
Now letting 
\begin{align}
  N,D            &\to\infty       \quad \text{with}\quad\frac{N}{\sqrt{D}} \equiv L=\text{const}\,,    \label{eq:limit}
\end{align}
Eq.~\eqref{eq:ring_intermediate_step} becomes
\begin{align}
  \partial_t u &= f(u) + q(R)\,\partial_{xx} u \,,\quad x\in[0,L] \nonumber 
  \intertext{where}
  q(R)         &=\sum_{j=1}^R j^2 = \frac 1 6 R(R+1)(2R+1). 
\end{align}
By rescaling $D\to q(R)D$, Eqs.~\eqref{eq:ring_system} finally become
\begin{subequations}\label{eq:fhn_1d_rd}
  \begin{align}
    \partial_t u &=u - \frac{u^3}{3} -v + \partial_{xx} u \label{eq:fhn_1d_rd_u}\\
    \partial_t v &= \ve (u+\beta)\,,                       \label{eq:fhn_1d_rd_v}
  \end{align}
\end{subequations}
with $x \in[0,L]$  and periodic boundary conditions $u(t,0)=u(t,L),\,v(t,0)=v(t,L)$, $L=\frac{N}{\sqrt{q(R)D}}$.

By the limit Eq.~\eqref{eq:limit}, the three parameters that determine the coupling and the topology of the ring network $N,\ R,$ and $D$ are translated into one parameter $L$, while the number of parameters of the local dynamics does not change.
Any constant in front of the second derivative can be set to unity by rescaling the spatial variable $x$.
Note that due to Eq.~\eqref{eq:limit}, decreasing $L$ in Eqs.~\eqref{eq:fhn_1d_rd} has the same effect as increasing $D$ in Eqs.~\eqref{eq:ring_system}.
Moreover, the information encoded in the parameter $R$ is lost when the limit is taken as in Eq.\eqref{eq:limit}.

With the parameter values being in the excitable regime  (as are the parameters chosen here, $\beta=1.1$, $\ve=0.04$), Eqs.~\eqref{eq:fhn_1d_rd} are known to support traveling-wave solutions.
These solutions move at constant speed $c$ and in a comoving frame they do not change their shape.
At  large domain sizes ($L>100$), Eqs.~\eqref{eq:fhn_1d_rd} have exactly two branches of traveling wave solutions, namely stable `fast waves' and unstable `slow waves'.
The domain size can become arbitrarily large without affecting the velocity and stability of these branches anymore.
When $L$ is decreased, these branches are connected, however, and together they form the dispersion relation $(L,\,c(L))$.
The dispersion relation is shown graphically in Fig.~\ref{fig:dispersion_relation_nw}(b) as the gray line, the branch of stable solutions is marked solid and the branch of unstable solutions is marked dotted.
At the chosen parameters of the local dynamics ($\ve=0.04,\,\beta=1.1$), the stable branch loses its stability at a critical domain size $L_\text{cr}=30.756$.
This can be understood as the result of the interaction of the traveling wave with its own tail due to the periodic boundary condition.
At this point, a torus bifurcation at which two conjugate complex points of the spectrum of the linearization of the traveling wave simultaneously cross the imaginary axis.
After this bifurcation, a complex series of secondary bifurcations finally leaves the branch of unstable waves with one point of the spectrum in the right half plane.
For more details about the bifurcations involved in the destabilization and the methods used to calculate the stability, we refer to the supplementary material and to \cite{KRU97,ROE07,BAC14}.

For the discrete ring system Eqs.~\eqref{eq:ring_system} with a given number of nodes $N$ and coupling range $R$, the critical domain size $L_\text{cr}$ determines an upper bound of the coupling $D_\text{high}=N^2/(q(R)L^2_\text{cr})$ for the stable propagation of waves by the limit Eq.~\eqref{eq:limit}.

\subsubsection{Traveling waves on the ring network}\label{sec:traveling-waves-ring-nw}
\begin{figure}[tb!]
\center
\includegraphics[width=\textwidth]{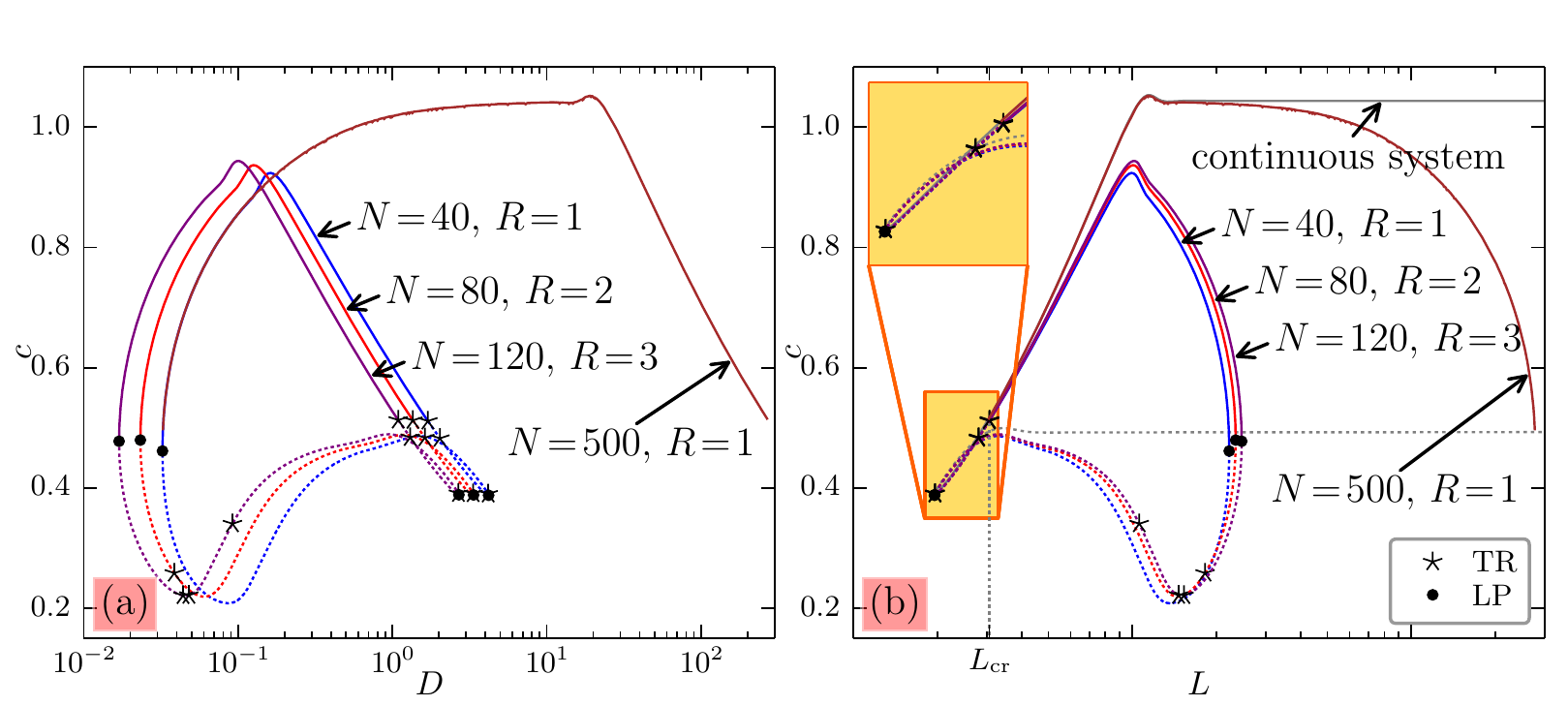}
\caption{Dispersion relations for traveling-wave solutions on the discrete ring network Eq.~\eqref{eq:ring_system} for different network sizes $N$ and nearest neighbor numbers $R$. 
(a) propagation speed $c$ vs.~coupling strength $D$. 
(b) propagation speed $c$ vs.~(virtual) domain size $L$. 
In (a) the limit points (saddle-node bifurcations) at $D_\text{low}$ for the same $R$ but different $N$ coincide. 
In (b) the destabilization points at $L_\text{cr}$ fall together for all networks. 
In (b) the dispersion relation for the continuous system Eq.~\eqref{eq:fhn_1d_rd} is shown in grey. 
The curve for $N=500$, $R=1$ has been obtained by numerical integration. 
All other curves have been obtained by path continuation.
The inset in (b) shows a blow-up of the yellow rectangle.
The filled dots denote limit points (LP), the asterisks denote torus bifurcations (TR).
Parameters: $\beta=1.1,\ \varepsilon=0.04$
\label{fig:dispersion_relation_nw}}
\end{figure}

In Eqs.~\eqref{eq:fhn_1d_rd}, $L$ can be set arbitrarily large without changing the shape, speed and stability of the traveling wave solutions.
This manifests itself in Fig.~\ref{fig:dispersion_relation_nw}(b) by the convergence of both stable and unstable branch of the dispersion relation of the continuum system (gray) to a finite propagation speed $c$.

However, in Eqs.~\eqref{eq:ring_system} increasing $L$ is equivalent to decreasing $D$ in Eqs.~\eqref{eq:ring_system}.
At low coupling strengths $D$, the discrete structure of the underlying ring network in Eqs.~\eqref{eq:ring_system} becomes important.
When the wave propagates in a discrete system, as Eq.~\eqref{eq:ring_system}, it cannot propagate at arbitrarily low coupling strengths \cite{BOO95,CAR05a}. 
The critical coupling strength $D$ at which the wave ceases to propagate has been approximately calculated in \cite{CAR05a}. 
We will point out what happens to the dispersion relation at $D_{\text{low}}$.
To the best of our knowledge this has not yet been investigated.

We calculate the dispersion relation from Eq.~\eqref{eq:ring_system} by choosing a certain number of nodes $N$ and performing a numerical continuation for the resulting full system of coupled ordinary differential equations, using AUTO-07p \cite{DOE09}.
Due to the closed ring topology, a traveling wave solution on the ring is given by a periodic orbit of the underlying $2N$ equations with period $T$.
The propagation speed $c$ can easily be calculated as $c=\frac{N}{T\sqrt{q(R)D}}$.
Using the transformed parameters $c$ and $L$ instead of $D$ and $T$, we can compare the dispersion relation of Eq.~\eqref{eq:ring_system} with that of Eq.~\eqref{eq:fhn_1d_rd} in the regime of high coupling strength $D$ or small (virtual) domain size $L$, respectively.
Dispersion relations obtained by continuation for $N=40,\,R=1$; $N=80,\,R=2$; and $N=120,\,R=3$ are displayed together with the branch of stable solutions obtained using numerical integration of Eqs.~\eqref{eq:ring_system} for $N=500,\,R=1$ in  the parameters $c$ vs. $D$ in Fig.~\ref{fig:dispersion_relation_nw}(a) and, using the transformed parameter $L=N/\sqrt{q(R)D}$ instead of $D$ in Fig.~\ref{fig:dispersion_relation_nw}(b).
Additionally, in Fig.~\ref{fig:dispersion_relation_nw}(b) the dispersion relation for the continuous system Eqs.~\eqref{eq:fhn_1d_rd} is shown.

For high coupling strengths, we find the same behavior as expected by the dispersion relation of Eq.~\eqref{eq:fhn_1d_rd}. 
However, for decreasing coupling strength, the propagation speed of the wave solutions does not stay constant, as it does in the continuous system. 
For the unstable `slow-waves' it increases whereas it decreases for the stable `fast-waves' until both branches meet at the critical low coupling strength $D_\text{low}$ in a saddle-node bifurcation. 
The coalescence of the branch of stable propagating waves and that of the unstable ones in a saddle-node bifurcation at the destabilization point has to our knowledge not yet been reported.
Because of this saddle-node bifurcation, the dispersion relation of wave solutions on the discrete ring Eq.~\eqref{eq:ring_system} is given by a closed curve.
The lower boundary of the coupling strength $D_\text{low}$ is a genuine effect of the discreteness of Eq.~\eqref{eq:ring_system} and does not depend on $N$ but it does depend on $R$.

In Fig.~\ref{fig:dispersion_relation_nw}(b), it is clearly visible that the smaller $L$ (i.e. the larger $D$), the closer the propagation velocity $c$ of the discrete system is to that of the continuum system.
It depends on the size $N$ and the coupling range $R$ of the network, above which value $L$ (or below which value $D$) the propagation velocity $c$ starts to deviate significantly.
For $N=500,\,R=1$, e.g., this is the case for $L\approx 200$ ($D\approx6.25$).
The reason for the deviation is that at low coupling strengths in the discrete system Eqs.~\eqref{eq:ring_system}, the excitation `hops' from node to node.
This mode of propagation is called saltatory propagation and is slower than the continuum-like propagation at higher coupling strengths \cite{CAR05a}.
The reason for saltatory propagation and for saltatory propagation being slower is that due to the low coupling strength the triggering of an excitation needs longer.
The actual transition from rest to excitation in contrast is fast and thus one node reaches full excitation before the next one starts the transition.

We summarize that on the ring network Eqs.~\eqref{eq:ring_system}, there is a lower bound of the coupling strength $D_\text{low}$ for stable wave propagation which depends only on the coupling range $R$ and not on the network size $N$.
The upper bound of the coupling strength for stable wave propagation $D_\text{high}$ in contrast is dependent on both $R$ and $N$ and is connected to the critical length for stable wave propagation $L_\text{cr}$ (see Sec.~\ref{sec:trav-waves-cont-lim} by $D_\text{high}=N^2/(q(R)L^2_\text{cr})$.
The dispersion relation for traveling waves on a ring network is given by a closed curve, where the branch of stable and the branch of unstable solutions meet twice.


\section{Wave-like solutions on small-world networks}\label{sec:wave_on_sw}
\subsection{Setup}
The network topology for our model is chosen to be a ring topology consisting of $N$ nodes, where each node is coupled to its $R$ neighbors to the left and its $R$ neighbors to the right, so that every node has degree $2R$. 
We perturb the topology by \emph{adding} a certain number $n$ of long-range links, where both ends of these links are chosen randomly \cite{NEW03}. This is a small modification of the well-known Watts-Strogatz small-world model \cite{WAT98}, where the long-range links are replacing links of the regular ring network. 
For a certain range of $n$ the Newman-Watts model shows small-world behavior as well, i.e., short average path length and high clustering coefficient, see Fig.~\ref{fig:nw_model_and_phaseplane}(a).

We are interested in the behavior of traveling waves on the ring, when the network topology is altered this way.
To this end, we numerically simulate a traveling-wave solution of Eq.~\eqref{eq:model} on a ring network without additional links.
At a certain instant of time, we instantly add a number $n$ of links at random positions to the network and monitor the resulting behavior.
The system can show two possible behaviors, it can either decay to the homogeneous steady state or it can show ongoing activity in the form of perturbed traveling waves.

In order to describe the collective effect in dependence on the number of additional links $n$, we consider the ensemble of all Newman-Watts small-world networks parametrized by network size $N$, coupling range $R$, and number of additional links $n$.
We define $\fsust(n,D ; N,R )$ as the fraction of realizations of a Newman-Watts small-world network that support sustained wave activity of Eqs.~\eqref{eq:model} with coupling strength $D$.
From now on $\fsust$ is the quantity we will be concerned with for the remainder of this work.

We determine $\fsust$ numerically by considering an ensemble of 200 Newman-Watts small-world networks for every combination $(N,R,n)$.
For each examined value of the coupling strength $D$, we integrate the dynamics with initial conditions as explained above for every element of this ensemble.
The fraction of realizations in this ensemble that support sustained wave activity gives our estimate of $\fsust(n,D ; N,R)$.
The number of additional links is varied from $n=1$ to $n=NR$.
When $n=NR$ there are as many additional links as links on the original ring and we have not observed a single case in which there was sustained wave activity for such a high number of additional links.
Thus there is no need to raise $n$ any further.
Also, in order to keep the computational effort feasible, we did not use every number $n$ between 1 and $NR$.
Instead we used about 50 values of $n$, distributed logarithmically between $1$ and $NR$.
This is also justified by the fact that for large $n$ the outcome for different, but close-by, $n$ hardly differs.
In the following, we omit $N$ and  $R$ from the argument of $\fsust$, writing just $\fsust(n,D)$, as $N$ and $R$ will be clear from the context and fixed when varying $n$ and $D$.
We do this for the full range of coupling strengths that support stable traveling wave solutions of the ring network  $D\in(D_\text{low},D_\text{high})$ (see Sec.~\ref{sec:traveling-waves-ring-nw}) and for rings of different sizes $N$ and different nearest neighbor numbers $R$. 
For the numerics we use a Runge-Kutta-Fehlberg method with adaptive timestep for fast simulation and reliability, the precision of the method is set to $10^{-4}$. 
We simulate for 6000 time units, assuming that the wave is stable if no decay to the stable homogeneous steady state has occurred within this time\footnote{Taking all solutions that actually decay within this timespan, the median time until decay is typically around 10 and always below 21. The 0.99-quantile is typically around 600 with one outlier ($N=1000$, $R=4$) where the 0.99-quantile is 3033. Thus it is safe to assume that almost all solutions that have ``survived'' the 6000 time units threshold will survive further.}.

\begin{figure}
\subfloat{%
  \includegraphics[width=0.5\textwidth]{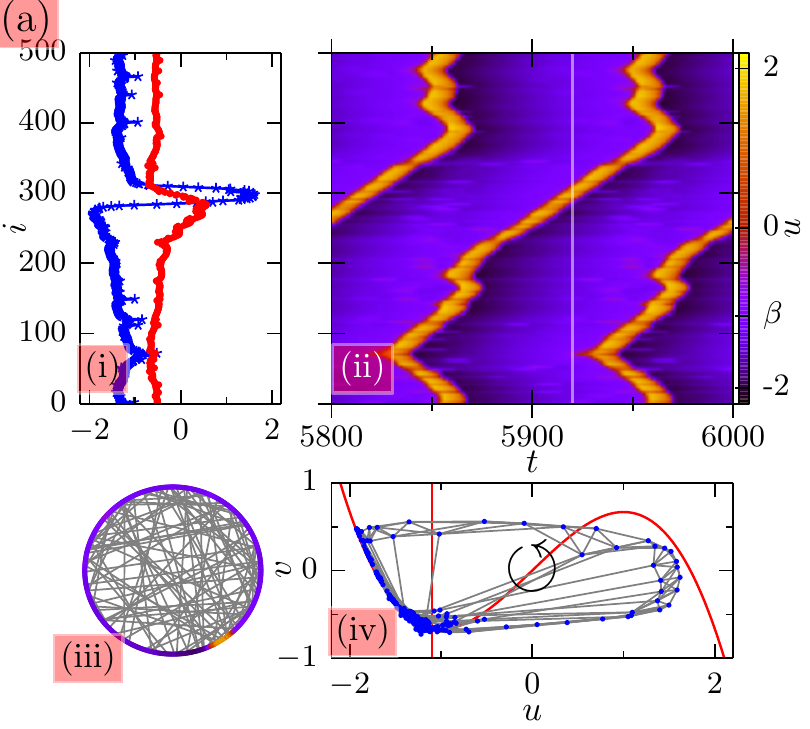}}%
\subfloat{%
  \includegraphics[width=0.5\textwidth]{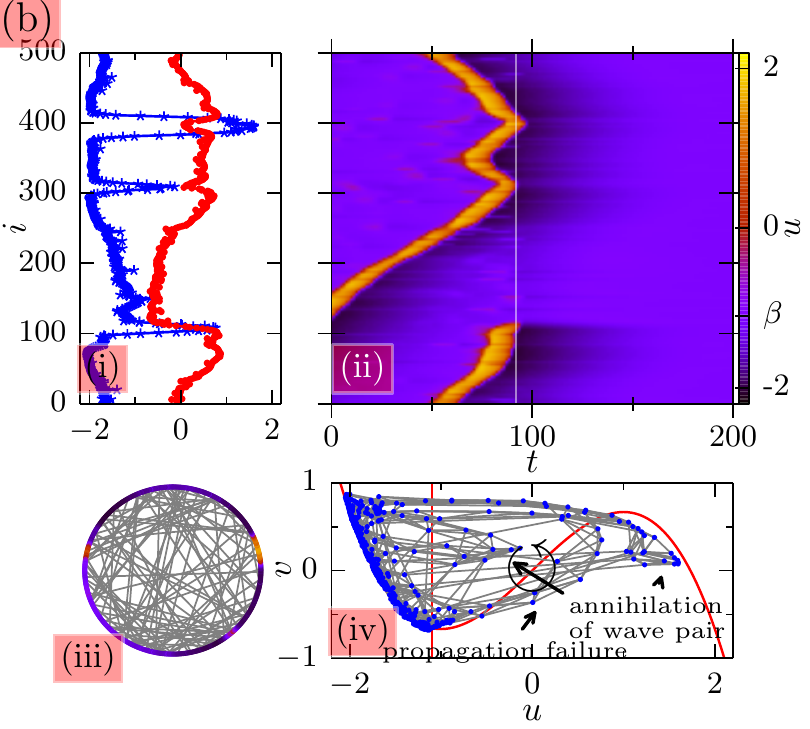}}%
\caption{Traveling wave solution on a Newman-Watts small-world network with  $N=500,\,R=3,\,n=102$.
            The coupling strength is $D=0.4$.
            (a) Sustained wave activity and
            (b) Propagation failure.
            The panels show
            (i)   Snapshot of the solution, $u_i$ (blue asterisks), $v_i$ (red dots) vs. node index $i$.
            (ii)  Space-time plot of activator variable $u_i$ (color-coded ), vertical white line marks snapshot in panel(i).
            (iii) Scheme of the network. Nodes are color coded according to their activator level at the time of snapshot (i).
                  Indices ascending counter-clockwise with $i=1$ at the top.
            (iv)  Phase portrait of all nodes in the $(u,v)$ plane (blue dots; snapshot), including all links of the network (gray lines) and nullclines (red).
                  Other parameters: $\ve=0.04,\,\beta=1.1$.
                  Animated versions of these Figures available under XXXX.}\label{fig:timeseries_500_102} 
\end{figure}

\subsection{Numerical observations}
\begin{figure}
\subfloat{%
  \includegraphics[width=0.5\textwidth]{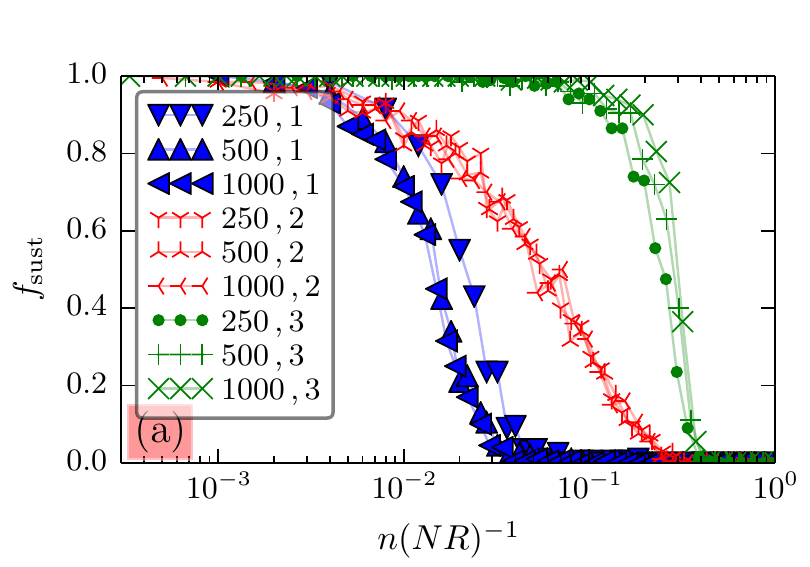}}%
\subfloat{%
  \includegraphics[width=0.5\textwidth]{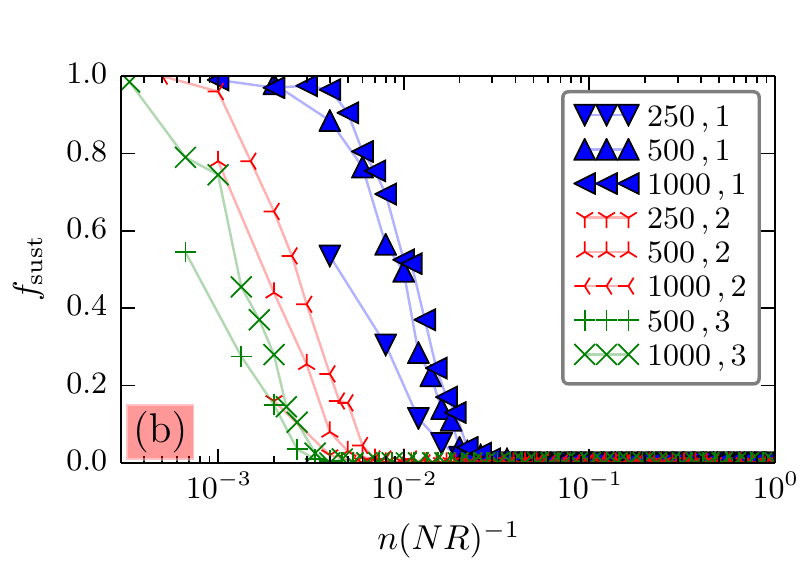}}
  \caption{ Fraction $f_\text{sust}$ of realizations (with $n$ additional long-range links) that support sustained activity of a traveling wave solution. 
  (a) $D=0.035$
  (b) $D=7.0$
Plots for ring networks of different sizes $N$ and coupling ranges $R$. The values ($N,\,R$) are given in the legend. 
Every data point has been calculated by simulating the dynamics on 200 realizations of the pertaining small-world network. 
Note that for (a) networks with the same $R$ show the same transition point, whereas in (b) the transition points of networks with the same $R$ so not coincide as well.
\label{fig:threshold_plot_over_n}
Parameters: $\beta=1.1,\ \varepsilon=0.04$
}%
\end{figure}

Generally we observe that the larger the number of additional links $n$, the more realizations do not support sustained wave activity.
Thus $\fsust$ always decreases with increasing $n$.
Finally, if $n$ becomes too large, no realizations with ongoing activity can be found anymore and $\fsust=0$.
We find that the transition takes place within a small range of $\frac{n}{NR}$, see Fig.~\ref{fig:threshold_plot_over_n}.
However, the point of transition varies considerably depending on
\begin{inparaenum}[(i)]
  \item the network size $N$,
  \item the nearest neighbor number $R$, and
  \item the coupling strength $D$.
\end{inparaenum}
Note that these parameters can also be chosen such that already for $n=1$, $f_\text{sust}$ is considerably below 1.0 (e.g. $N=250$, $R=2$, $D=7.0$ in Fig.~\ref{fig:threshold_plot_over_n}(b), leftmost point for those parameters).

In general, it is not possible to pinpoint the exact mechanism that causes propagation failure when multiple additional links are present.
For example there are different realizations with the same $n$ that for the same coupling strength $D$ either support sustained wave activity or do not, see for example Fig.~\ref{fig:timeseries_500_102}. 
On the other hand, we can also find one realization that for one value of $D$ supports ongoing activity and for a slightly different $D$ does not.
In Fig.~\ref{fig:timeseries_500_102} it can also be seen that propagation failure is not caused by collision with another counterpropagating wave.
We have only observed pairwise generation of secondary waves, see, e.g., (white) vertical line in Fig.~\ref{fig:timeseries_500_102}(b).
As the mutual annihiliation of counterpropagating waves also takes place only pairwise (see vertical line in Fig.~\ref{fig:timeseries_500_102}(b,ii)), this mechanism cannot change the number of simultaneously occurring waves from even to odd or vice versa and therefore cannot cause propagation failure.
The example in Fig.~\ref{fig:timeseries_500_102}(b) is selected such that at the time of the snapshot, mutual annihilation of wave pairs and propagation failure occur simultaneously.
In Fig.~\ref{fig:timeseries_500_102}(b,i), the nodes' states at the time of the snapshot are displayed and those groups of nodes which undergo propagation failure or annihiliation of a wave pair are marked by arrows in Fig.~\ref{fig:timeseries_500_102}(b,iv).
It can be seen in this plot that in decaying to the left branch of the $u$-nullcline, the group of nodes undergoing propagation failure sweeps over a large section of the middle part of the $u$-nullcline.
The group of nodes that experience the mutual annihilation of a wave pair, in contrast, cross the middle part of the $u$-nullcline at a small corridor.

\begin{figure}[tb!]
  \subfloat{%
    \includegraphics[width=0.5\textwidth]{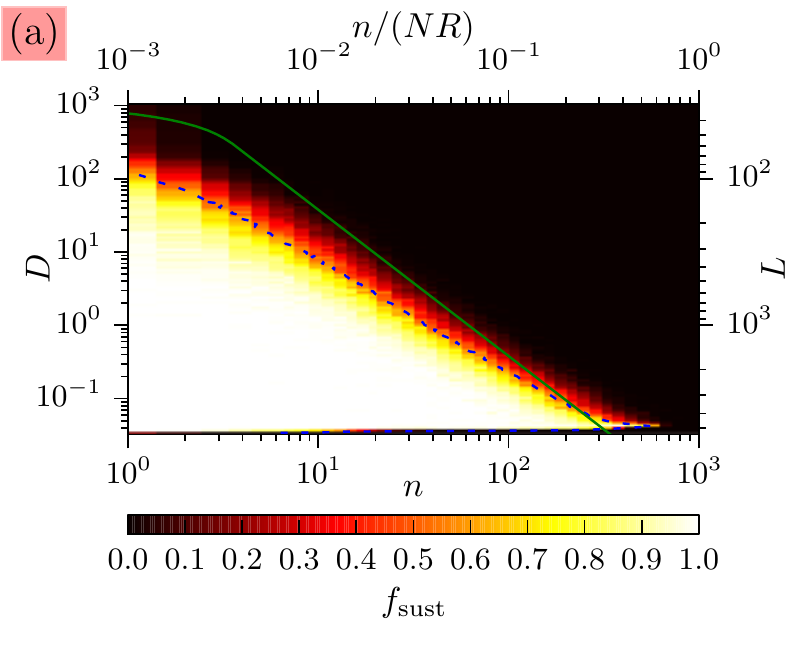}}%
  \subfloat{%
    \includegraphics[width=0.5\textwidth]{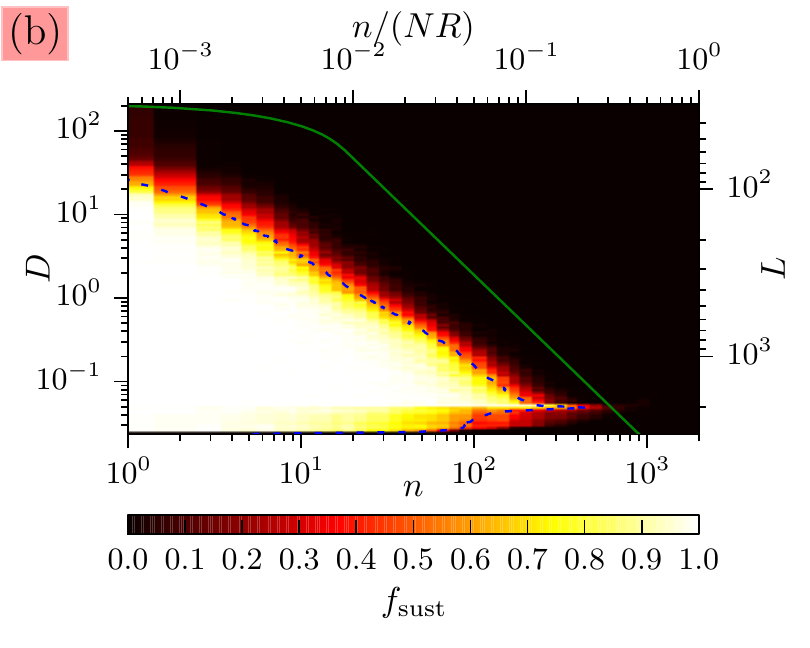}}%

  \subfloat{%
    \includegraphics[width=0.5\textwidth]{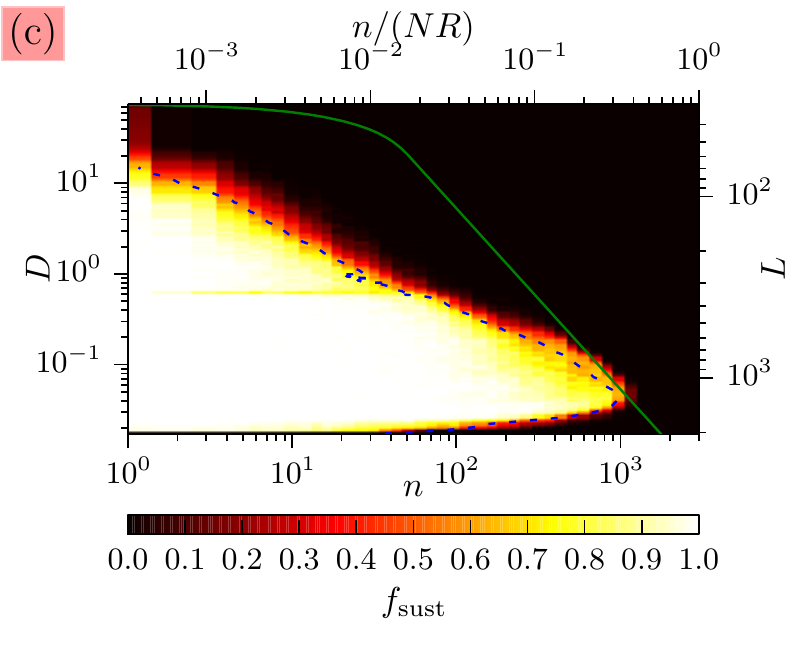}}%
  \subfloat{%
    \includegraphics[width=0.5\textwidth]{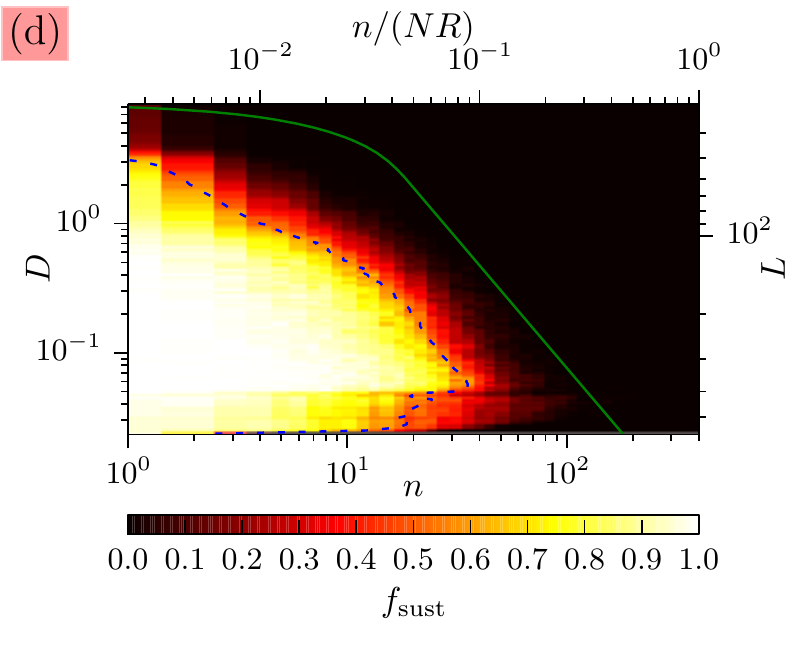}}%
  \caption{Density plots of fraction of sustained wave activity $f_\text{sust}$ (color coded) in the $(n,D)$-plane.
           (a) $N=1000$ and $R=1$,
           (b) $N=1000$ and $R=2$,
           (c) $N=1000$ and $R=3$,
           (d) $N=200$ and $R=2$.
           The green line gives the location of the destabilizing bifurcation of Eq.~\eqref{eq:fhn_1d_rd_mf}, transformed back to $n$ and $D$ (see Sec.~\ref{sec:continuum_limit}). 
           Note that the transitions in the discrete regime in (b) and (d) take place at the same values of $D$, whereas the ones in the continuous regime take place at the same values of $L$. Parameters: $\beta=1.1,\ \varepsilon=0.04$. \label{fig:surv_plot_sim_1000}}
\end{figure}

For a systematic investigation, we examine the fraction $\fsust(n,D)$ of realizations that support an ongoing activity.
In Fig.~\ref{fig:surv_plot_sim_1000}, color density plots of $\fsust(n,D)$ are shown for different $N$ and $R$.
We numerically calculate the isolines $(n_{0.5},D_{0.5})$.
These are defined by $\fsust(n_{0.5},D_{0.5})=0.5$ and shown as blue dotted lines in Fig.~\ref{fig:surv_plot_sim_1000}.

When $D$ approaches either $D_\text{low}$ or $D_\text{high}$, while fixing the the number of additional links, $\fsust$ is approaching zero.
This is seen as the transition to black at the bottom and top of Figs.~\ref{fig:surv_plot_sim_1000}.
This behavior is expected from the results in Sec.~\ref{sec:traveling-waves-ring-nw}, as there are no stable traveling-wave solutions beyond these values even on the unperturbed ring.
However, as the propagation of the unperturbed wave differs in the two regimes, so does the sustained propagation or the propagation failure on the small-world network.
As can be seen in Fig.~\ref{fig:threshold_plot_over_n}(a), when the coupling strength is low, the network size $N$ does not play a role and networks with the same $R$ show the same transition point in $n/(NR)$.
When the coupling strength is higher, as in Fig.~\ref{fig:threshold_plot_over_n}(b), the transition points do not coincide as well.

\begin{figure}
 \subfloat{%
   \includegraphics[width=0.5\textwidth]{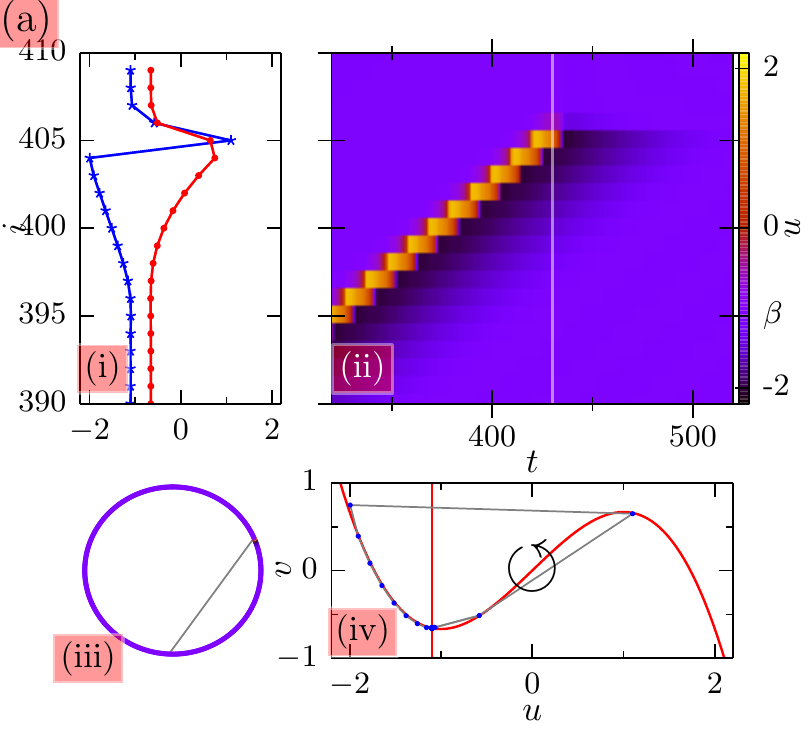}}%
 \subfloat{%
   \includegraphics[width=0.5\textwidth]{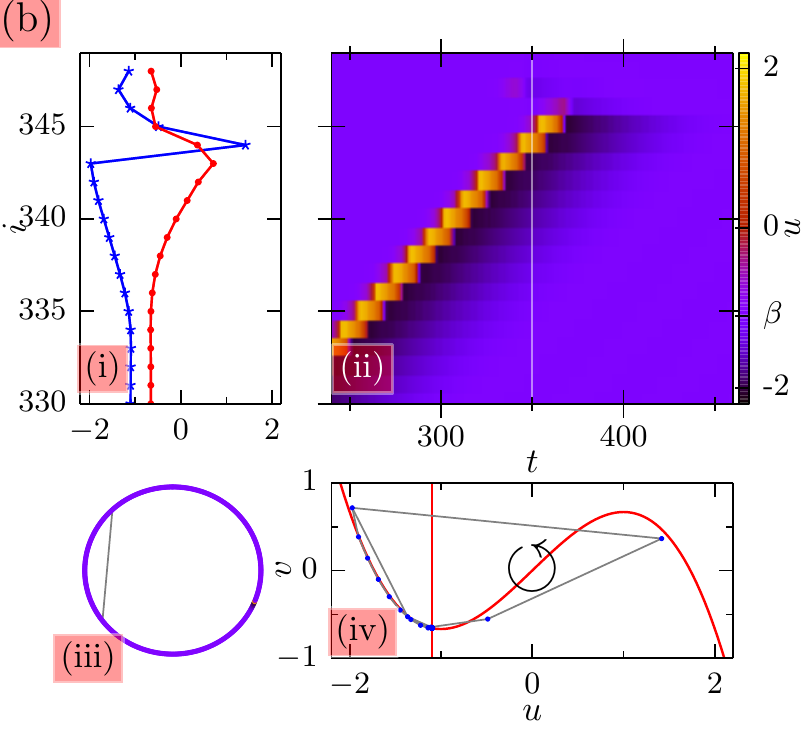}}%

 \subfloat{%
   \includegraphics[width=0.5\textwidth]{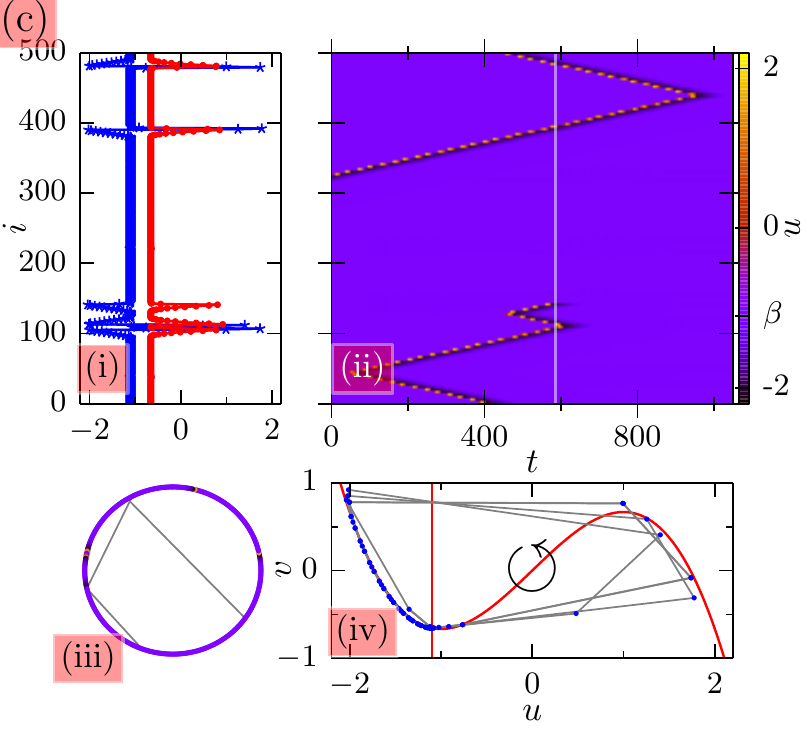}}%
 \subfloat{%
  \includegraphics[width=0.5\textwidth]{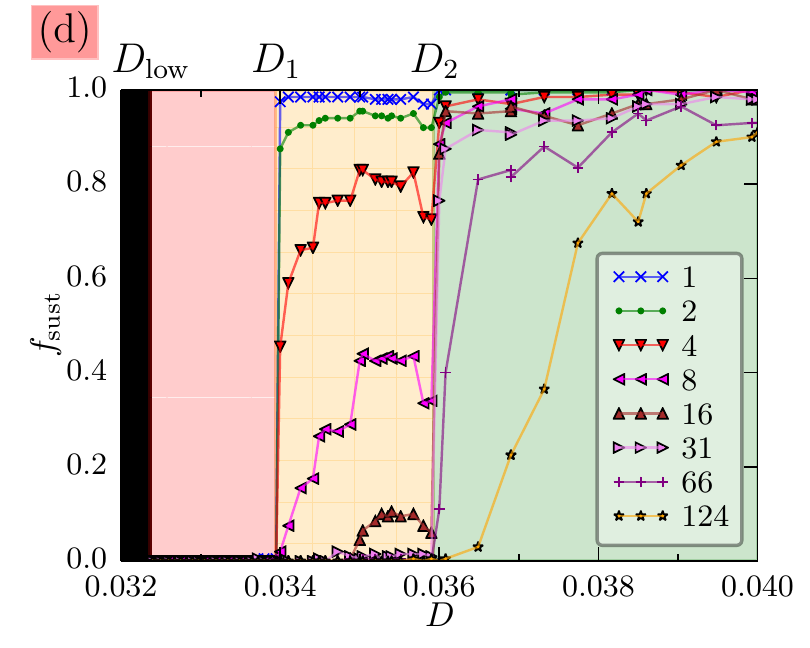}}%
   \caption{
Exemplary behavior of traveling-wave solutions in the regime of very low coupling strengths $D$. 
(a)-(c) Same as Fig.~\ref{fig:timeseries_500_102}, with 
(a) $D_\text{low}<D\approx 0.0332<D_1,\,n=1$, propagation suppression by backcoupling to an unexcited node and 
(b) $D_1<D\approx 0.0347<D_2,\,n=2$, propagation suppression by raised inhibitor level due to previous sub-threshold excitation mediated by additional link and 
(c) $D_2<D\approx 0.0373,\,n=4$, excitation of secondary wave pairs possible, propagation suppression by two additional links ending in node 142 and 
(d) Fraction of sustained wave activity $f_\text{sust}$ vs. coupling strength $D$ for the discrete limit, legend gives values of $n$. 
Parameters: $\beta=1.1,\ \varepsilon=0.04,\ N=500,\ R=1$. Animated versions available under XXXX  \label{fig:ts_discrete_limit}}%
\end{figure}

\subsubsection{Low $D$ --- discrete regime (Fig.~\ref{fig:ts_discrete_limit})}
In this regime of low $D$, the propagation of the excitation takes place in a saltatory fashion.
By this, we mean that the excitation `hops' from node to node.
More precisely, the state of one node reaches maximum excitation before the next node starts becoming excited.
This behavior can be seen clearly in the exemplary time series Fig.~\ref{fig:ts_discrete_limit}.

In addition to $D_\text{low}$, discussed in Sec.~\ref{sec:traveling-waves-ring-nw}, we find two more values $D_1,\,D_2$ with $D_\text{low} < D_1 < D_2 $ at which the effect which added links have changes suddenly.
These values are independent of $N$, but they depend on $R$.

If $D < D_\text{low}$, we find no stable traveling wave solutions whatsoever, even for $n=0$.
This is expected, as for $D<D_\text{low}$ there are no stable traveling solutions even on the ring without additional links (see Sec.~\ref{sec:traveling-waves-ring-nw} and Fig.~\ref{fig:dispersion_relation_nw}).

If $D_\text{low} < D < D_1$, one additional link will lead to propagation failure once the traveling wave reaches one of the nodes which this link joins.
When the node that is about to become excited has one end of the additional link, the other end will point to a node that is in the rest state.
This is because the propagation is saltatory, there is only one node that is excited at one instance of time.
In this range of $D$, this link is sufficient to prevent this node from becoming excited, and thus the propagation is quenched.
As a consequence, $\fsust(n>0,D)=0$ but $\fsust(n=0,D)=1$ in this range of $D$.
This behavior is illustrated in Fig.~\ref{fig:ts_discrete_limit}(a).
There the exemplary network has one additional link from node 246 to node 406. 
Propagation suppression by coupling back to an unexcited node happens as soon as the wave reaches node 406.

If $D_1 < D < D_2$, one additional link may lead to propagation failure, but it does not necessarily (Fig.~\ref{fig:ts_discrete_limit}(b)
In this range of $D$, a traveling-wave solution can pass one end of the additional link without being suppressed.
When the node at the first end of the additional link becomes excited, this excitation is also coupled to the node at the remote end.
Because of the coupling scheme of the ring, this node is coupled to more nodes in the rest state than the successor node to the node at the first end of the link.
As a consequence, the node at the first end will be able to trigger a full excitation in its successor node but not in the node at the remote end.
Here a sub-threshold excitation is generated, which does not propagate further.
However, this sub-threshold excitation leads to an increased inhibitor level which takes a certain time to decay back to the steady state value.
If the remote end of the shortcut will be reached by the traveling wave before the inhibitor level has sufficiently decayed, the propagation will stop.
This behavior leads to decrease of $\fsust(n,D)$ for increasing $n$ in the mentioned range of $D$, because as $n$ increases, the more likely it becomes that one of the additional links spans a short enough distance.
An exemplary time series for this behavior is shown in Fig.~\ref{fig:ts_discrete_limit}(b).
There the exemplary network has two additional links from node 60 to 176 and from 343 to 347.
Propagation failure happens by the raised inhibitor level at node 347 due to previous sub-threshold excitation mediated by the additional link.

If $D > D_2$, a traveling wave solution can still pass one end of the additional link without becoming suppressed.
But now a full excitation, leading to a pair of traveling waves with opposite propagation directions will be generated at the remote end of the additional link.
This can be seen very clearly in Fig.~\ref{fig:ts_discrete_limit}(c,ii), where secondary wave pairs are generated at node 43 and at node 127. 
As traveling waves are only generated pairwise and the annihilation also takes place in pairs, the mechanism for propagation failure must work differently. 
One possible mechanism that we found and that can also be seen in Fig.~\ref{fig:ts_discrete_limit}(c) is that two or more additional links end very close to each other (or even on the same node). Then, as can be seen in fig.~\ref{fig:ts_discrete_limit}(c) it might 
(i) not be possible to excite a secondary wave pair here and also 
(ii) propagation can stop here due to the strong coupling back to the rest state as in the regime below $D_1$. Generally, as $f_\text{sust}$ is close to unity for small $n$ in this regime, propagation failure seems to be mostly caused by more complex mechanisms which need several additional links.
In the exemplary timeseries for Fig.~\ref{fig:ts_discrete_limit}(c) there are four additional links from node 90 to 127, 38 to 142, 142 to 221 and 43 to 326.
The propagation stops at node 142, where two additional links end. 
A secondary wave pair is never excited at this node because the other additional link couples back this node to the rest state.
Up to the point of propagation failure

To summarize: For $D < D_\text{low}$, no traveling wave solutions exist for any $n\ge 0$. 
For $D_\text{low} < D < D_1$ no stable traveling wave solutions exist for $n>0$. 
For $D_1 < D < D_2$ $f_\text{sust}$ decreases with increasing $n$, no secondary waves can be excited. 
For $D > D_2$ the excitation of secondary waves is possible, leading to a sudden rise of $f_\text{sust}$ for intermediate $n$.

\begin{table}
\begin{tabular}{r||l|l|l}
  $R$ &  $D_\text{low}$ & $D_1$         & $D_2$  \\
\hline
1     &  $\approx0.0324$    & $\approx0.0339$    & $\approx0.0359$   \\
2     &  $\approx0.0233$    & $\approx0.0235$    & $\approx0.0481$  \\
3     &  $\approx0.0169$    & $\approx0.0170$    & $\approx0.5890$   \\
\end{tabular}
\caption{Approximate transition values $D_\text{low},\,D_1$ and $D_2$ for different nearest neighbor numbers $R$ in the discrete limit.\label{tab:kappa_vals}}
\end{table}

The observed approximate values for $D_\text{low},\,D_1$ and $D_2$ are given in Table~\ref{tab:kappa_vals}.
We note that the values for $D_\text{low}$ and $D_1$ (i) decrease with increasing coupling range $R$ as does (ii) the distance between the two. 
This is expected as (i) increasing $R$ has a similar effect as raising $D$ on the ring and (ii) as the long-range links have the same weight as the local links, they have less impact, if $R$ becomes larger.
$D_2$ on the contrary increases with increasing $R$. This is also expected for the same reason as (ii).
For $R=3$, $D_2$ is not even located in the regime of saltatory propagation anymore.

\begin{figure}
   \centering
   \includegraphics[width=0.5\textwidth]{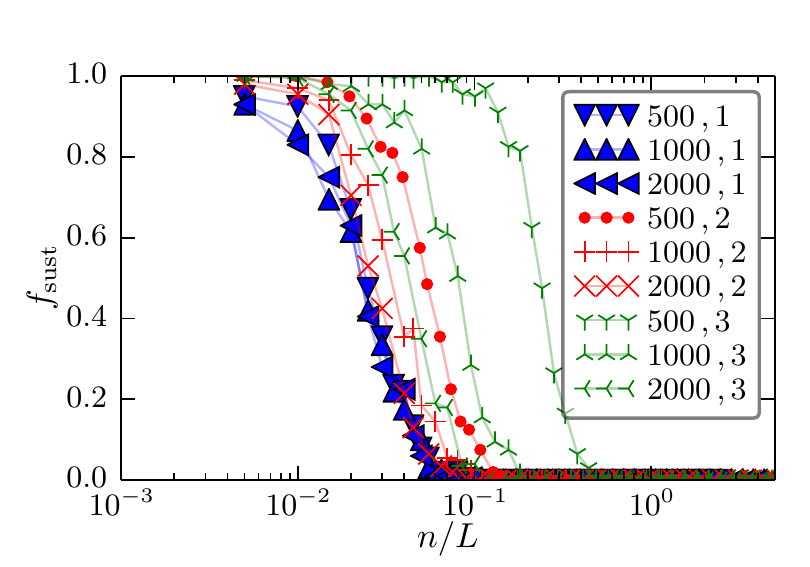}
   \caption{Fraction of sustained wave activity $f_\text{sust}$ in the continuum regime of $D$. 
            $n/L$ at $L\!=\!200$, where $L=\frac{N}{\sqrt{q(R)D}}$, legend gives values $(N,R)$.
           Other parameters: $\beta=1.1,\ \varepsilon=0.04$ \label{fig:threshold_continuum_regime}}%
\end{figure}

\subsubsection{High $D$ --- continuum regime (Fig.~\ref{fig:threshold_continuum_regime})}
At high coupling strengths $D$, we do not find distinct values of $D$ at which the overall behavior of $\fsust$ changes drastically as in the discrete regime. 
We observe the excitation of secondary waves in the entire continuum regime.
Again, as the generation of secondary waves occurs only pairwise, this mechanism can not lead to the decay of all activity directly by pairwise annihilation of counterpropagating waves.

We find as the main mechanism for propagation failure again the distribution of too many ends of the additional links in a small region, thus coupling back nodes that are in the excited state too strongly to nodes having a low-activator concentration. This effect can be seen very nicely in the $(u,v)$ diagrams in Fig.~\ref{fig:timeseries_500_102}(b), where all nodes that constitute the original wave are pulled over the middle-part of the $u$-nullcline.

We also observe another notable effect:
For nearest neighbor numbers higher than $R=1$, there is a fixed coupling strength (depending on $R$ but not on $N$) at which $f_\text{sust}$ starts to decrease already at lower numbers $n$ of additional links. This can be seen very well in Fig.~\ref{fig:surv_plot_sim_1000}(c) at $D\approx 1$. So far we have no explanation for the mechanism behind this phenomenon.

If we are in the regime of high coupling strength $D$, $f_\text{sust}$ decreases with increasing $D$ until $D$ reaches the maximum value $D_\text{high}$ above which no traveling wave solutions are found even for $n=0$ (ring without additional links), see Sec.~\ref{sec:traveling-waves-ring-nw}.
Along these lines it turns out that the parameters $L=\frac{N}{\sqrt{q(R)D}}$ and $\sigma \equiv \frac{2n}{L}$ are better suited to describe the behavior of $\fsust$ independently of the network size $N$ and (almost) independently of the nearest neighbor number $R$. See Sec.\ref{sec:analytic} for a derivation and discussion.

This is shown in Fig.~\ref{fig:threshold_continuum_regime}, where the fraction of sustained wave activity $f_\text{sust}$ is plotted versus $n/L$ for various $N$ and $R$.
The coupling strength $D$ in this plot is adapted for each ($N,\,R$) so that $L=200$ is constant in that figure.
Note that the transition points from sustained wave activity to propagation suppression coincide better, the lower $R$ and the larger $N$.
For higher $R$ ($R=2,3$) the networks with smaller $N$ need to have (much) smaller $D$ to have the same $L$ and thus are not well located in the continuum regime anymore.
Thus in order to show that the approximation works very well in that case, networks with $N>2000$ would need to be simulated, which would have been numerically too expensive within the scope of this work.

\section{Analytic mean-field approximation}\label{sec:analytic}
\subsection{High $D$ --- continuum limit}\label{sec:continuum_limit}
In order to include the effect of the long-range links into the continuum limit description of Sec.~\ref{sec:trav-waves-cont-lim}, we split the adjacency matrix $\mathcal{A}_{ij}$ in Eq.~\eqref{eq:model} into two parts. $\mathcal{A}_{ij} \equiv \mathcal{R}_{ij} + \mathcal{S}_{ij}$, with $\mathcal{R}_{ij}$ being all links of the original ring network and $\mathcal{S}_{ij}$ being the additional randomly added links. 
Thus the dynamics reads
\begin{align}
  \dot u_i  &=  f(u_i,v_i) + D\left( \sum_{j=1}^N \mathcal{R}_{ij}(u_j-u_i) + \sum_{j=1}^N \mathcal{S}_{ij}(u_j-u_i) \right) \nonumber \\
            &=  f(u_i,v_i) + D \sum_{j=1}^R(u_{i+j}+u_{i-j}-2u_i)  + D\sum_{j=1}^N \mathcal{S}_{ij}(u_j-u_i)   \nonumber \\
            &=  f(u_i,v_i) + \frac{\tilde{D}}{q(R)} \sum_{j=1}^R(u_{i+j}+u_{i-j}-2u_i)  + \frac{\tilde{D}}{q(R)}\sum_{j=1}^N S_{ij}(u_j-u_i)\nonumber
\end{align}
where in the last equality a rescaling $D\to\tilde{D}=D q(R)$ (see Sec.~\ref{sec:traveling-waves-ring-nw}) has been used, and the tilde will be dropped in the following.

The ring part of the coupling can be treated in the same way as in Sec.~\ref{sec:trav-waves-cont-lim}. For the small-world part of the coupling, we assume a large number of additional links and distribute the entries in $\mathcal{S}_{ij}$ equally over all entries of the entire matrix $\mathcal{S}_{ij}$ which leaves $\mathcal{S}_{ij}=\frac{2n}{N^2}$ a constant. 
For easier readability, we consider only the small-world part of the coupling term $s_i$:
\begin{align}
  s_i  &\equiv \frac{D}{q(R)}\sum_{j=1}^N \mathcal{S}_{ij}(u_j-u_i) \nonumber \\
       &=  \frac{1}{q(R)}\frac{N^2}{L^2}\sum_{j=1}^N\left(\frac{2n}{N^2}(u_j-u_i)\right) \nonumber \\
       &=  \frac{1}{q(R)}\frac{2n}{L^2}\sum_{j=1}^N(u_j-u_i) \nonumber
\intertext{In performing the transition to the continuum description, we replace the sum $\sum_{j=1}^N$ by the integral $\int_{0}^Ldy$
and introduce the mean value $\bar u$}
 s(x)&= \frac{1}{q(R)}\frac{2n}{L^2} \left(\int_{0}^L u(y) dy  - \int_{0}^Lu(x)dy \right) \nonumber \\
       &=   \frac{1}{q(R)}\frac{2n}{L}   \left(\bar u -u(x) \right) \nonumber
\end{align}
The continuum limit including the additional long-range links reads
\begin{subequations}
\label{eq:fhn_1d_rd_mf}
\begin{align}
  \partial_t u &= \frac{u^3}{3}-u -v + \partial_{xx} u + \sigma (\bar u -u) \label{eq:fhn_1d_rd_mf_u} \\
  \partial_t v &= \ve (u+\beta)                                            \label{eq:fhn_1d_rd_mf_v} \\
  x            &\in[0,L]\text{ and }(u,v)(t,0)=(u,v)(t,L)\,,              \nonumber
\end{align}
\end{subequations}
with $L=\frac{N}{\sqrt{q(R)D}}$ and $\sigma=\frac{2n}{q(R)L}$.

With this mean-field approximation, the four coupling parameters ($N,\ R,\ D$ and $n$) reduce to two parameters ($L$ and $\sigma$).
This kind of global feedback coupling has also been studied for the Rinzel-Keller model in \cite{HEM98}.

\subsection{Approximate boundary of wave propagation}

\begin{figure}
   \includegraphics[width=\textwidth]{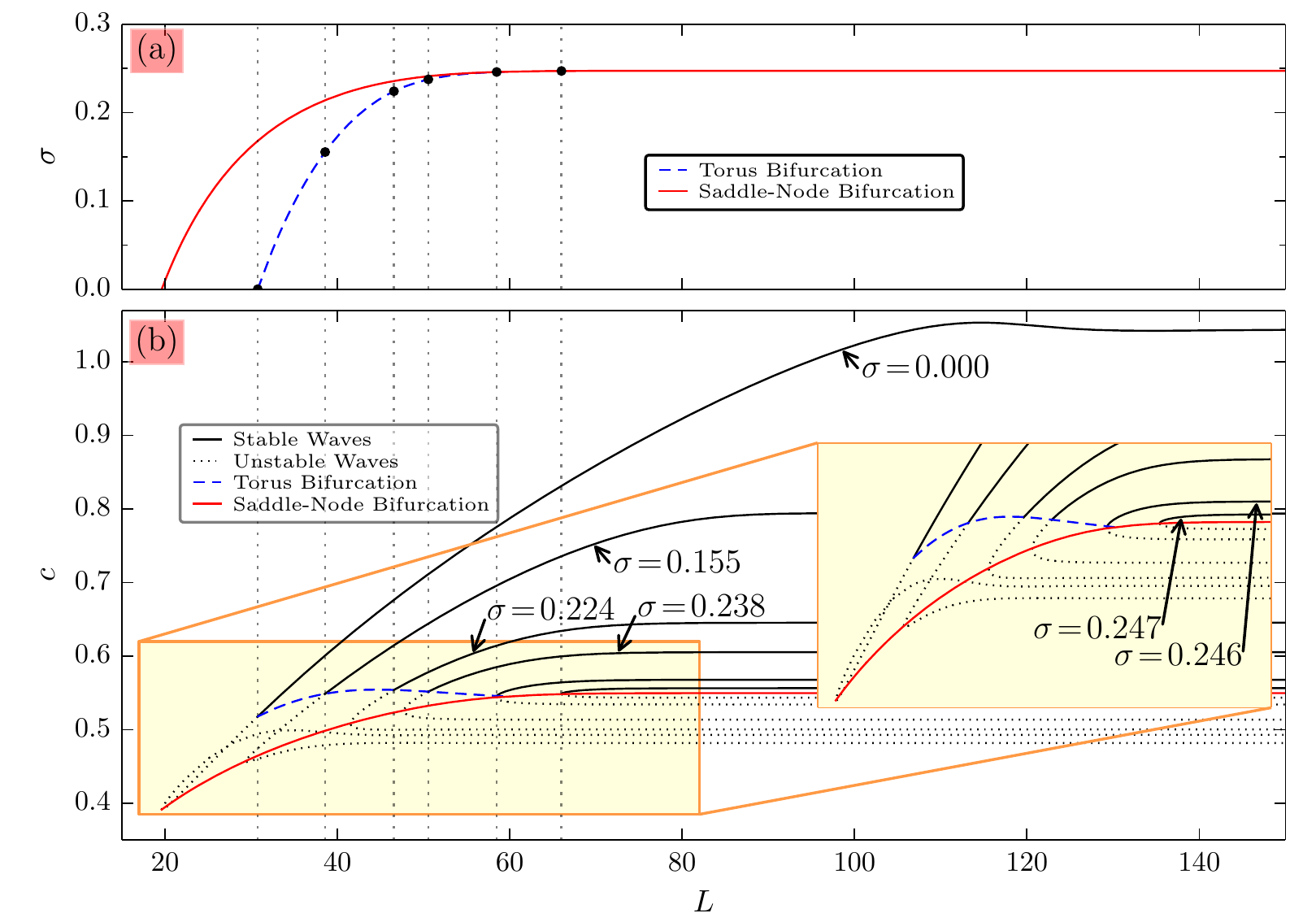}
   \caption{Dispersion relation for the mean-field approximation Eq.~(\ref{eq:fhn_1d_rd_mf}). 
            (a) Curve of the destabilizing torus bifurcation at $L_{\text{cr}}$ (blue dashed) and of the saddle-node bifurcation (red solid) in $(L,\sigma)$ space. 
            Black dots indicate the $(L, \sigma)$ values of the destabilization points in the dispersion relations shown in (b).
            (b) Propagation velocity $c$ vs $L$: branches of stable (black solid) and unstable (black dotted) traveling waves for different values of mean-field coupling strength $\sigma$, 
            curves of destabilizing torus bifurcation (blue dashed), and curve of the saddle-node bifurcation (red  solid) in $(L,c)$ space. 
            The inset shows a blow-up of the yellow rectangle.
            Parameters: $\beta=1.1,\ \varepsilon=0.04$.\label{fig:dispersion_relation_mean_field} }
\end{figure}

If $\sigma=0$, Eqs.~\eqref{eq:fhn_1d_rd_mf} are the same as Eqs.~\eqref{eq:fhn_1d_rd}. 
Employing the same methods as in Sec.~\ref{sec:trav-waves-cont-lim}, we examine the change of the dispersion relation $c(L)$ if $\sigma$ is increased, see Fig.~\ref{fig:dispersion_relation_mean_field}.

If $\sigma=0$, stable wave propagation is possible down to a minimum value of  $L_\text{cr}\approx 30.756$ (see Sec.~\ref{sec:trav-waves-cont-lim}).
If $\sigma$ is increased, $L_\text{cr}$ increases as well, i.e., the parameter range of $L$ for stable propagation becomes smaller(Fig.~\ref{fig:dispersion_relation_mean_field}(b)). 
However, $L_\text{cr}$ goes to infinity when $\sigma$ approaches $\sigma_\text{max}\approx 0.247$ from below, so that above $\sigma_\text{max}$, no stable propagation is possible at all.

Note that at $\sigma\approx 0.246$, the mechanism of destabilization changes, when the destabilizing torus bifurcation coincides with a saddle-node bifurcation (limit point).
The destabilizing torus bifurcation is indicated by blue dashed lines in Fig.~\ref{fig:dispersion_relation_mean_field} and the destabilizing saddle-node bifurcation is indicated by a red solid line.

We display the loci of destabilization as a curve $L_\text{cr}(\sigma)$ in Fig.~\ref{fig:dispersion_relation_mean_field}(a).
$L_{cr}(\sigma)$ in (a) is connected with the instability points of the dispersion relations shown in Fig.~\ref{fig:dispersion_relation_mean_field}(b) as indicated by the vertical dotted lines.

This curve can be transformed to a curve $n_0(N,R,D)$, yielding an approximation for the boundary in $n$ above which no realizations of a small-world network will support stable traveling waves. 
It is shown as the green solid line in the heatmap plots of $\fsust$ Figs.~\ref{fig:surv_plot_sim_1000}.
The transition in $n$ to quenched wave activity happens at lower values of $n$. 
This is expected, as a significant contribution by the coupling term arising through the long-range links can only occur if the difference in activator concentration at both ends of the link is large.
This is only the case if the node at one end of the shortcut is in the excited state (wave peak).
Thus the critical link density is only important in part of the network. 
Of course, in a random network this is more likely to occur in an (arbitrary) part than in the entire network.
Also note that the approximation becomes worse for higher $R$. 
Moreover, there is always an optimum coupling strength $D$ where $n$ can be highest without disturbing the propagation of the wave.
This optimum $D$ is a result of the transition between the discrete and the continuum regime.

\section{Conclusion}\label{sec:conclusion}
We have studied the propagation of a solitary pulse (or wave) on a ring network and the influence of small-world perturbations of the topology upon the propagation.
Already on the unperturbed ring topology, there are two regimes.
One regime corresponds to high coupling strength, in which the behavior of the system resembles that of a continuous reaction-diffusion system with mean-field coupling.
The other regime is associated with low coupling strength, in which the discrete nature of the network is important and the behavior differs from that of a reaction-diffusion system.

In each regime, a too large number of long-range links leads to failure of wave propagation. 
However, the mechanisms which lead to the suppression of the traveling wave differ in the two regimes.

We have identified three different subregimes of coupling strength in the weak-coupling regime, which are sharply separated from each other.
In the first one (lowest coupling strength), one additional link, regardless of the distance it spans, is enough to prevent propagation. 
In the second one, one additional link can be sufficient to prevent propagation if the distance it spans is not too large. 
For coupling strengths above the second subregime, secondary wave pairs can be created through the long-range links.
For the latter coupling strengths, the mechanism for the quenching of a traveling wave is similar to that in the continuum regime.

In the strong-coupling regime, the main mechanism appears to be a too large number of additional links in the excited part of the wave (high activator concentration).
Those links collectively ``pull'' the excited part back over the threshold trajectory of the system and thus lead to propagation failure.
We have successfully approximated this behavior in the continuum limit by including a mean-field coupling term in the equations of the continuous reaction-diffusion system.

\section*{Acknowledgement}
This work was supported by DFG in the framework of SFB 910. Helpful discussions with Philipp H\"{o}vel, Judith Lehnert and Niklas 
H\"{u}bel are acknowledged.

\bibliographystyle{unsrt}

\end{document}


\title{Supplementary material for 
''Effect of small-world topology on wave propagation on networks of excitable elements''}
\author{T. Isele and E. Sch\"oll\\
Institut f\"ur Theoretische Physik, \\Technische Universit\"at Berlin, \\10623 Berlin, Germany
\vspace{0.5em}\\
\texttt{schoell@physik.tu-berlin.de} }
\maketitle
\newpage

\section{Excitability in the FitzHugh-Nagumo system}

The local dynamics of Eqs.~\eqref{main:eq:model} are given by
\begin{subequations}
\label{eq:local_dyn}
  \begin{align}
    \dot u &= u- \frac{u^3}{3} -v \\
    \dot v &= \ve(u-\beta)\,.
  \end{align}
\end{subequations}
The periodic orbit which emerges from the Hopf bifurcations at $\abs\beta=1$ shows a canard explosion \cite{BEN81a} immediately after the bifurcation point ($\abs\beta\lesssim 1$), evolving to the large-amplitude limit cycle of a relaxation oscillator. 
In the regime where $\abs\beta>1$ and no limit cycle (stable or unstable) exists in the local dynamics, Eqs.~\eqref{eq:local_dyn} serve as a paradigmatic example for type-II excitability. 

If we view the Eqs.~\eqref{eq:local_dyn} as a singularly perturbed system, we note that by Fenichel theory \cite{FEN79}, the normally hyperbolic parts of the $u$-nullcline (which give the critical manifold of the system in the limit $\ve\to 0$) are perturbed to attracting and repelling \emph{slow manifolds} $S_\ve^a,\ S_\ve^r$, respectively.
Any trajectory that passes near $S_\ve^r$ is repelled from it with $v$ staying almost constant (because $\ve \ll 1$) until it hits $S_\ve^a$.
A trajectory that follows the repelling slow manifold $S_\ve^r$ for a `considerable amount of time' is called a canard trajectory \cite{BEN81a}.
In the case of the FitzHugh-Nagumo dynamics it is a common choice to take such a canard trajectory as the threshold trajectory, e.g., one uses the trajectory that goes through the point where the repelling and attracting parts of the critical manifold meet (which is given by the $u$-nullcline), see Fig.~\ref{main:fig:nw_model_and_phaseplane}(b).

\section{Dispersion relation of traveling waves in the continuous system}
\begin{figure}
\center
\includegraphics[width=\textwidth]{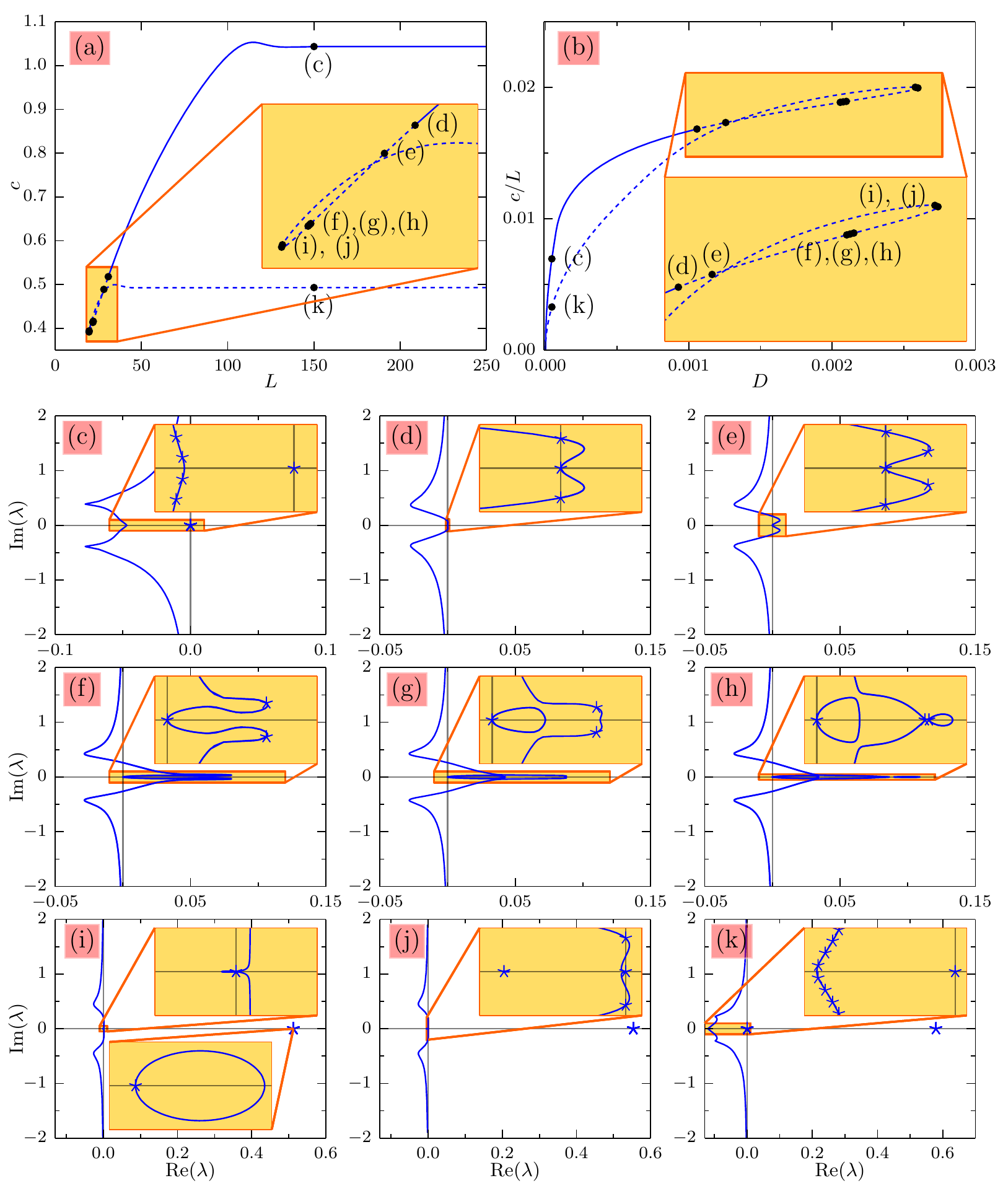}
\caption{Dispersion relation for wave trains (or traveling pulses subject to periodic boundary conditions on a domain of size $L$.). 
(a) Dispersion relation of Eq.~\eqref{main:eq:fhn_1d_rd} i.e., propagation velocity $c$~vs.~$L$. 
(b) Same dispersion relation in the transformed coordinates $D=1/L^2$ and $c/L$ which are more adequate for networks. 
(c)-(k) Leading part of the essential spectrum for selected points marked in panels (a),(b). 
The insets in (a), (b) show an enlarged view of the yellow rectangles, the insets in (c)-(k) show a blow-up of the spectrum near the origin.
Those points that are in the spectrum for the system with {\em periodic} boundary conditions are marked by an asterisk.
Parameters: $\beta=1.1,\ \varepsilon=0.04$
\label{fig:dispersion_relation_cont}}
\end{figure}

With the parameter values being in the excitable regime  (as are the parameters chosen here, $\beta=1.1$, $\ve=0.04$), Eqs.~\eqref{main:eq:fhn_1d_rd} are known to support traveling-wave solutions.
These solutions move at constant speed $c$ and in a comoving frame they do not change their shape.
At a given $L$ there can be different traveling wave solutions traveling at different speeds, namely stable `fast waves' and unstable `slow waves'.
When varying $L$, these branches are connected, however, and together they form the dispersion relation $(L,\,c(L))$.
For more details, we refer to \cite{KRU97,ROE07,BAC14}.

We calculate the dispersion relation of traveling wave solutions of Eqs.~\eqref{main:eq:fhn_1d_rd} by transforming to the comoving frame $\xi=x-ct$ and doing numerical branch continuation using AUTO-07p \cite{DOE09}.
The equations in the comoving frame (and written as a first-order system) are called the profile equations and read
\begin{subequations}
\label{eq:profile_eqs}
  \begin{align}
    \dot u &= w \\
    \dot v &= -\frac{\ve}{c} (u-\beta) \\
    \dot w &= -cw - \left( u- \frac{u^3}{3} -v \right)\,.
  \end{align}
\end{subequations}
Due to the periodic boundary conditions, a traveling wave solution of Eqs.~\eqref{main:eq:fhn_1d_rd} appears as a periodic orbit in Eqs.~\eqref{eq:profile_eqs}.
The domain size $L$ defines the period of this orbit and the propagation velocity $c$ is an additional parameter that needs to be solved for during the continuation.

The spectrum of the linear stability analysis around such a traveling wave solution is very closely connected to the spectrum around a wave-train in an infinitely extended spatial domain subject to the same differential equation.
The spectrum around this wave-train consists only of continuous components which together are called the essential spectrum \cite{RAD07b}.
By linearizing Eqs.~\eqref{main:eq:fhn_1d_rd} around a traveling wave solution using a Bloch-expansion ansatz and transforming to the comoving frame, the essential spectrum can be calculated by numerical continuation of the resulting boundary value problem using the Bloch wavenumber as continuation parameter.
This boundary value problem reads
\begin{subequations}
\label{eq:fhn_wave_spectrum_bvp}
  \begin{align}
    \delta u '   &=  \delta w                                                                            \label{eq:fhn_wave_spectrum_bvp_u}   \\
    \delta v '   &=  \frac 1 c \left( \lambda \delta v - \ve \delta u \right)                            \label{eq:fhn_wave_spectrum_bvp_v}   \\
    \delta w '   &=  \lambda \delta u - \left[ (1-u_\text{tw}^2)\delta u - \delta v \right] - c \delta w  \label{eq:fhn_wave_spectrum_bvp_w}   \\
    \delta u(L)  &= \e{i2\pi\nu} \delta u(0)                                                             \label{eq:fhn_wave_spectrum_bvp_bcu} \\
    \delta v(L)  &= \e{i2\pi\nu} \delta v(0)                                                             \label{eq:fhn_wave_spectrum_bvp_bcv} \\
    \delta w(L)  &= \e{i2\pi\nu} \delta w(0) \,,                                                         \label{eq:fhn_wave_spectrum_bvp_bcw}
  \end{align}
\end{subequations}
where $u_\text{tw}$ is (the $u$ part of) the profile of a traveling wave solution to Eqs.~\eqref{eq:profile_eqs} with spatial period $L$ and propagation speed $c$, $\nu$ is the Bloch wavenumber in units of $L/(2\pi)$, and $\lambda\in\mathbb{C}$ is an element of the spectrum.

We have used this method which was proposed in \cite{RAD07b} to calculate the leading part of the essential spectrum of the linearization of Eq.~\eqref{main:eq:fhn_1d_rd} around the connected wave train.
The essential spectrum is continuous and contains every point of the spectrum of a traveling wave solution of Eq.~\eqref{main:eq:fhn_1d_rd} subject to periodic boundary conditions.
These points additionally fulfil the condition of an integer Bloch-wavenumber $\nu$.
The method we use to calculate the spectrum is elaborated in great detail in \cite{RAD07b} including hints on the implementation in AUTO.

We have calculated the dispersion relation for a traveling wave solution of Eqs.~\eqref{main:eq:fhn_1d_rd} (subject to periodic boundary conditions) as well as the essential spectrum of the related wave-train at selected solutions and those points of the essential spectrum that are in the spectrum of the traveling wave subject to periodic boundary conditions.
The results are shown in Fig.~\ref{fig:dispersion_relation_cont}.
For the dispersion relation of traveling wave solutions of Eqs.~\eqref{main:eq:fhn_1d_rd}, we report the following:
Starting with a  stable traveling-wave solutions and large $L$, the calculated spectrum tells us that a destabilization of the wave occurs at a critical value of $L_{\text{cr}}\approx 30.756$ by two complex conjugate eigenvalues with nonzero imaginary part.
However, this point does not yet mark the lowest possible propagation speed $c$. 
This is attained by the branch of unstable solutions after lowering $L$ further to a value of $L\approx19.6$. 
After that, the branch of unstable solutions continues in form of a fold bifurcation in direction of rising $L$ with the speed converging against $c\approx0.493$. 
On this branch, $L$ can increase without bounds, and this branch is associated with the transition to an (unstable) solitary traveling pulse. 
Also on the stable branch, $L$ can increase without bounds but the solutions do not lose stability and converge to a stable propagating solitary pulse.

After the destabilization at $L_{\text{cr}}$, the spectrum changes in a complex way, leaving in the end an isolated eigenvalue as the only object in the right half-plane.
The dispersion relation (propagation velocity $c$ vs. domain size $L$) including the leading parts of the (essential) spectrum for selected points (c)-(k) is illustrated in Fig.~\ref{fig:dispersion_relation_cont}(a). 
In the pictures of the spectrum, the essential spectrum is marked by continuous lines and the values of the essential spectrum that occur for periodic boundary conditions are marked with asterisks.
The latter are the relevant ones for our model.
The order of occurrence of the points (c)-(k) is the following, 
\begin{inparaenum}[(a)]
 \item [(c)] All eigenvalues are in the left half-plane (except for the one that is always present at zero corresponding to the Goldstone mode of translation invariance),
 \item [(d)] two complex conjugate eigenvalues crossing the imaginary axis at $L_{\text{cr}}$,
 \item [(e)] a second pair of complex conjugate eigenvalues crossing the imaginary axis,
 \item [(f),(g)] from the leading part of the essential spectrum, a circle comprising the Goldstone mode eigenvalue is forming and detaching,
 \item [(h)] the first two eigenvalues that have crossed the imaginary axis merge on the real axis and split in different directions,
 \item [(i)] one of the eigenvalues that has merged on the real axis crosses zero,
 \item [(j)] the second two eigenvalues that have crossed the imaginary axis cross the imaginary axis again in the opposite direction.
 \item [(k)] One eigenvalue is left in the right half-plane, the rest of the eigenvalues are in the left half-plane (again except for the Goldstone mode eigenvalue at zero).
\end{inparaenum}
Thus after the first instability at $L_{\text{cr}}$, a scenario with several secondary instabilities emerges.

As discussed in the main text, the reaction-diffusion system Eq.~\eqref{main:eq:fhn_1d_rd} is a good approximation for the ring network Eq.~\eqref{main:eq:ring_system} for large $N$ and large $D$.
In Fig.~\ref{fig:dispersion_relation_cont}(b), we have displayed the dispersion relation of Fig.~\ref{fig:dispersion_relation_cont}(a) in transformed coordinates $c/L$ vs. $D=1/L^2$.
In these coordinates it is easier to compare this dispersion relation of a continuum system to that of the discrete ring system Eqs.~\eqref{main:eq:ring_system}. 
On the ring network, the above sequence of changes in the spectrum would translate to the following sequence of bifurcations:
A torus bifurcation through which the wave loses its stability at (d), another torus bifurcation at (e), a saddle-node bifurcation at (i), a torus bifurcation at (j). 
The minimum domain size $L_\text{cr}$ for which a traveling-wave solution is stable translates to a maximum coupling strength according to Eq.~\eqref{main:eq:limit}.
We denote this maximum coupling strength by $D_\text{high}=\frac{N^2}{q(R)L_\text{cr}^2}$.

\bibliographystyle{unsrt}